\documentclass{elsart}
\usepackage{amsmath}
\usepackage{amssymb}
\usepackage{graphicx}
\usepackage{dcolumn}

\newcommand{\bcc}{\ensuremath{\mathbf{C}}}

\newcommand{\bx}{\ensuremath{\mathbf{x}}}

\newcommand{\bu}{\ensuremath{\mathbf{u}}}

\newcommand{\uu}{\mbox{\boldmath$u$}}

\newcommand{\br}{\ensuremath{\mathbf{r}}}

\begin{document}
\begin{frontmatter}
\title
{Kinetic Theory of Turbulence Modeling:\\
Smallness Parameter, Scaling and \\Microscopic Derivation of Smagorinsky Model}
\author{Santosh Ansumali }
\address{ETH-Z\"urich, Department of Materials and Department of Energy Technology,
CH-8092 Z\"urich, Switzerland}
\author{ Iliya V.\ Karlin}
\address{ETH-Z\"urich, Department of Materials, 
CH-8092 Z\"urich, Switzerland\\
Institute of Computational Modeling RAS, 660036 Krasnoyarsk, Russia}
\author{Sauro  Succi}
\address{
Istituto Applicazioni Calcolo, CNR, viale Policnico 137, 00161  Roma, Italy}
\begin{abstract}
A mean-field approach (filtering out subgrid scales) is applied to 
the Boltzmann
equation in order to derive a subgrid turbulence model  
based on kinetic theory. 
It is demonstrated
that the only Smagorinsky type  model which survives 
in the hydrodynamic limit on the
viscosity time scale is  the  so-called tensor-diffusivity model.
Scaling of the filter-width with Reynolds number and Knudsen number
is established. This sets the first rigorous step in deriving turbulence
models from kinetic theory.
\end{abstract}
\end{frontmatter}
\section{Introduction}
\label{intro}

  The application of  a filtering procedure
to equations of hydrodynamics (Navier-Stokes equations) in order
to construct a subgrid model is often used for the turbulence modeling
\cite{Pope2000}. The aim of such  models is to  take into account the 
effects of
subgrid scales as an extra stress term in the
hydrodynamic equations for the resolved scale fields. Further, the
subgrid scale terms should be representable in terms of the resolved
fields.  This procedure, like any other attempt to coarse-grain the
Navier-Stokes equations, runs into  the closure problem due to the
nonlinearity of the equation and due to the absence of scale separation.
On the other hand, in  statistical
physics, good schemes to obtain closure approximations are
known for nonlinear evolution equations (with a well-defined
separation of scales). Unfortunately,  attempts to borrow such
schemes fail for the Navier-Stokes equations. 
The fundamental reason for this failure of the coarse-graining
procedures on the
Navier-Stokes equations is the absence of scale separation. Further, 
 the length over which the  equation is coarse-grained (the filter width in
the present case) is completely arbitrary (and in practice dictated by
the available computational resources), and  cannot be justified a priori on
physical grounds. 

In this paper, we  show that a coarse-grained description of
hydrodynamics using  the microscopic theories is possible. Specifically,  
we apply 
the standard filtering procedure (isotropic Gaussian  filter)
not on the Navier-Stokes equations but on the Boltzmann kinetic equation.
We recall that the Navier-Stokes equations are a well defined limit of the
Boltzmann equation (the hydrodynamic limit), whereas 
{\it the filtering operation
and going to the hydrodynamic limit are two distinct operations which do not
commute},
because kinetic fluctuations generally do not annihilate upon filtering.
 By doing so, we obtain the following results:

\begin{itemize}
\item {\it Smallness parameter:} The smallness parameter of the present theory
is the usual kinetic-theory Knudsen number ${\rm Kn}$,
\begin{equation}
\label{Knudsen}
{\rm Kn}
 =  \frac{ \nu}{L c_{\rm s}}\sim\frac{{\rm Ma}}{{\rm Re}},
\end{equation}
where 
${\rm Ma}$ is the Mach number and  ${\rm Re}$ is  the Reynolds number,  $\nu$
is the kinematic viscosity, $c_{\rm s}$ is the sound speed and  $L$ is the
characteristic macroscopic length.  Smallness of ${\rm Kn}$
rules emergence of {\it both}, the usual viscosity terms, and the subgrid contributions,
on the viscosity time scale of the filtered Boltzmann equation (that is, in the first-order
Chapman-Enskog solution to the filtered Boltzmann equation). 
\item {\it Scaling:} In the coarse-grained
representation obtained by filtering, the filter-width $\Delta$   (for the Gaussian filter,
$\Delta^2$ is proportional to the covariance) is the  smallest
length-scale to be resolved. 
The  requirement that contributions from the
subgrid scales appear in the kinetic picture at the time scale of
molecular relaxation time (viscosity time scale)
sets the scaling of $\Delta$ with the Knudsen number as follows:
\begin{equation}
\Delta=kL\sqrt{{\rm Kn}},
\label{Scaling1}
\end{equation}
where $k$ is a nonuniversal constant  which  scales neither with $L$,
nor with ${\rm Kn}$. For the sake of simplicity, we set $k=1$ in
all the further computations. 
Equations (\ref{Knudsen}) and (\ref{Scaling1}) imply that  the 
filter-width scales with the Reynolds number  as follows:
\begin{equation}
\label{Scaling2}
\Delta\sim {\rm Re}^{-1/2}.
\end{equation}
While the Kolmogorov length, $l_{\rm K}$, scales as $l_{\rm K}\sim {\rm Re}^{-3/4}$, 
we have
\begin{equation}
\label{Scaling3}
\frac{\Delta}{l_{\rm K}}\sim {\rm Re}^{1/4}.
\end{equation}
Thus, the filtering scale is larger than the Kolmogorov scale when
${\rm Re}$ is large enough.

\item {\it Subgrid model:} With the above smallness parameter (\ref{Knudsen}),
 and the scaling  (\ref{Scaling1}), we rigorously derive the following subgrid
pressure tensor $P^{\rm SG}_{\alpha\beta}$,   
in addition to the usual (advection and viscosity) terms
in the momentum equation:

\begin{eqnarray}
P^{\rm SG}_{\alpha\beta}  &= & \frac{{\rm Kn}  L^2  \overline{\rho}}{12} \left[ 
\overline{S}_{\alpha\gamma} -\overline{\Omega}_{\alpha\gamma}
\right]\left[ 
\overline{S}_{\gamma\beta} +\overline{\Omega}_{\gamma\beta}
\right]\nonumber\\ &=&  \frac{\nu \overline{\rho} \, c_{\rm s} \, L}{12 } \left[ 
\overline{S}_{\alpha\gamma} -\overline{\Omega}_{\alpha\gamma}
\right]\left[ 
\overline{S}_{\gamma\beta} +\overline{\Omega}_{\gamma\beta}
\right].
\label{Smago1}
\end{eqnarray}
Here $\overline{\rho}$ is the filtered density, and summation
convention in spatial components is adopted. For any function $X$,
$\overline{X}$ denotes the filtered value of $X$. 
Furthermore, the filtered  rate of the strain tensor
$\overline{S}_{\alpha\beta}$ and the filtered rate of the rotation tensor
$\overline{\Omega}_{\alpha\beta}$ depends only the large scale velocity,
$\overline{ u_{\alpha}}$:
\begin{eqnarray}
\begin{aligned}
\overline{\Omega}_{\alpha\beta} &= \frac{1}{2}\left\{ \partial_{\alpha}
\overline{ u_{\beta}} -
    \partial_{\beta}\overline{ u_{\alpha}} \right\},\\
\overline{S}_{\alpha\beta} &= \frac{1}{2}\left\{ \partial_{\alpha} \overline{ u_{\beta}} +
    \partial_{\beta} \overline{ u_{\alpha}}\right\}.
\end{aligned}
\end{eqnarray}
The derived subgrid model belongs to the class  of  Smagorinsky models \cite{Smagorinsky},
 and the
tensorial structure of the subgrid pressure tensor (\ref{Smago1}) corresponds
to the so-called tensor-diffusivity subgrid  model (TDSG)
introduced by Leonard  \cite{Leonard74}, and
which became popular after the work of Clark et al \cite{Clark79}. 
Here, it is interesting to recall that in the class of existing
Smagorinsky models the TDSG  is one of only a few  models in which the sub-grid
scale stress tensor remains
 frame-invariant 
 under arbitrary time-dependent rotations  of  reference
 frame \cite{Jap}. Furthermore, the TDSG model belongs to a subclass of
 Smagorinsky models which take into account the backscattering of energy from the
 small scale to the large scales \cite{Pope2000}. Beginning with the seminal work of
 Kraichnan \cite{Kraich}, importance  of the backscattering of energy
in turbulence modeling  is commonly recognized.

\item {\it Uniqueness:} The result (\ref{Smago1}) requires only  isotropy
of the filter but otherwise is independent of the particular functional form
of the filter. There are no other subgrid models 
different from (\ref{Smago1}) which can be derived from kinetic theory
by the one-step filtering procedure.
In other words, higher-order spatial derivatives are not neglected
in an uncontrolled fashion, rather, they are of the order ${\rm Kn}^2$, and thus
do not show up on the viscosity time scale.

\item {\it Nonarbitrary filter-width:} Unlike the phenomenological TDSGM  where
the prefactor in Eq.\ (\ref{Smago1}) remains an unspecified ``$\Delta^2$'',
kinetic theory  suggests that the filter-width cannot be set at will,
rather, it  must respect the physical parameterization 
(specific values of ${\rm Re}$, ${\rm Kn}$  etc)
of a given setup when the subgrid model is used for numerical simulation.
Recent findings that  simulations of the TDSG  model become  unstable for
 large $\Delta$ \cite{Jap2}
 is in qualitative agreement with the present result that the
 filter-width  $\Delta$
 cannot be made arbitrary
 large. 

\end{itemize}

The structure of the paper is as follows: In section \ref{KT}
we set up the kinetic theory for the subsequent coarse-graining and 
derivation of the subgrid model.
It is important to stress that the only requirement on the choice of the
kinetic equation in the present context is that it gives the 
Navier-Stokes hydrodynamic
equations in the appropriate fully resolved  limit. 
For that reason we choose to work with
a recently introduced minimal kinetic model \cite{KFOe99,AKOe2003} 
which is sufficient
for our present purpose.
Filtered kinetic equation is obtained in section \ref{FKT}.
In  section \ref{CE} we derive the subgrid model (\ref{Smago1}) using the
Chapman-Enskog method \cite{Chapman} for the filtered kinetic model.
This derivation uniquely defines the scaling (\ref{Scaling2}) from
the requirement that the subgrid terms appear on the viscous time scale.
Finally, results are summarized, and some directions of future
research are discussed in section \ref{END}.


\section{Kinetic theory}
\label{KT}
For the present discussion, the particular  choice  of the  kinetic model 
is unimportant as long as the hydrodynamic limit of the kinetic theory
is the usual Navier-Stokes equations at least up to the order 
$O({\rm Ma^3})$.  
We demonstrate the whole procedure in detail for a recently introduced minimal
discrete-velocity kinetic model \cite{KFOe99,AKOe2003}. 
As the final result is just the same in  two and three dimensional
cases, for the sake of simplicity we chose to
demonstrate the whole procedure using  the two-dimensional model ($D=2$). 
The kinetic equation is,

\begin{equation} 
\label{LBM} 
\partial_t f_i+ C_{i\alpha} \partial_{\alpha}f_i = 
-\tau^{-1} \left(f_i- f_i^{\rm eq}  \right), 
\end{equation} 
where $f_i(\bx,t)$, $i=1,\dots 9$ are populations of 
discrete velocities $\bcc_i$:

\begin{align}
C_x &= \left\{0, 1,0,-1,0,1,-1,-1, 1\right\},  \\
C_y &=  \left\{0, 0,1,0,-1,1,1,-1, -1\right\}.
\end{align}

The local equilibrium $f_i^{\rm eq}$ is the conditional minimizer of the 
 the  entropy  function $H$: 
\begin{equation} 
\label{app:H} 
H=\sum_{i=1}^{9} f_{i}\ln\left(\frac{f_{i}}{W_i} \right),
\end{equation}
 under the constraint of local conservation laws: 
\begin{equation}
 \sum_{i=1}^{9} f^{\rm eq}_i \{ 1,\  \bcc_{i}\}.
=\{\rho,\  \rho \bu   \}.
\end{equation}
The weights $W_i$ in the equation (\ref{app:H}) are:
\begin{equation}
W = \frac{1}{36} \left \{16, 4,4,4,4,1,1,1,1 \right \}.
\end{equation}
The explicit  expression  for  $f^{\rm eq}_i$ reads:
\begin{align}
\label{TED}
\begin{split}
f^{\rm eq}_i =\rho W_i\prod_{\alpha=1}^{D} 
\left(2 -\sqrt{1+ 3 {u_{\alpha}}^2}\right)
\left(
\frac{ 2 u_{\alpha}+ 
\sqrt{1+ 3 {u_{\alpha}}^2}}{1- u_{\alpha}}
\right)^{C_{i \alpha}}.
\end{split}
\end{align}

Below, it will prove convenient to work in
the moment representation rather than in the population representation.
Let us 
choose the following orthogonal set of basis vectors in 
the $9$-dimensional phase space of
the kinetic equation (\ref{LBM}):
\begin{align}
\begin{split}
\psi_1 &=  \left\{1, 1,1,1,1,1,1,1, 1\right\},  \\
\psi_2 &=  \left\{0, 1,0,-1,0,1,-1,-1, 1\right\},  \\
\psi_3 &=  \left\{0, 0,1,0,-1,1,1,-1, -1\right\},  \\
\psi_4 &=  \left\{0, 0,0,0,0,1,-1,1, -1\right\},  \\
\psi_5 &=  \left\{0,1,-1,1, -1, 0,0,0,0\right\},  \\ 
\psi_6 &= \left\{0, -2,0,2,0,1,-1,-1, 1 \right\},  \\
\psi_7 &=  \left\{0, 0,-2,0,2,1,1,-1, -1\right\},  \\
\psi_8 &=  \left\{4, -5,-5,-5,-5,4,4,4, 4\right\},  \\
\psi_9 &=  \left\{4, 0,0,0,0,-1,-1,-1, -1\right\}.  \\
\end{split}
\end{align}
The orthogonality of the chosen basis is in the sense of the usual
Euclidean scalar product, i.e., 
\begin{equation}
\sum_{k=1}^{9} \psi_{ik} \psi_{kj} = d_i\delta_{ij},
\end{equation}
where $d_i$ are some constants needed for the normalization (the basis
vectors are orthogonal but not orthonormal). 
We define new variables  $M_i$, $i=1,\dots,9$ as:
\begin{equation}
M_i = \sum_{j=1}^{9} \psi_{ij} f_j,
\end{equation}
where $\psi_{ij}$ denotes $j$th component of the  $9$-dimensional vector $\psi_i$.
Basic hydrodynamic fields are $M_1=\rho$, $M_2=\rho u_x$, and
$M_3=\rho u_y$. The remaining  six moments are related to higher order moments
of the distribution (the pressure tensor $P_{\alpha \beta}= \sum f_i
C_{i \alpha} C_{i \beta}$ and the third order moment $Q_{\alpha \beta
  \gamma }= \sum f_i
C_{i \alpha} C_{i \beta} C_{i \gamma}$ and so on),  as: $M_4 =
P_{xy}$, 

\begin{eqnarray}
\begin{aligned}
M_5 &= P_{xx} - P_{yy},\\
 M_6  &= 3  \sum f_i C_{iy}^2 C_{ix} - 2 M_2,\\
  M_7  &= 3 \sum f_i C_{ix}^2 C_{iy}-2 M_3  
\end{aligned}
\end{eqnarray}  
The explicit form of the stress tensor in term of the new set of
variables  is:
\begin{align}
\begin{split}
P_{xy} &= M_4,\\
P_{xx}&=  \frac{2}{3} M_1 + \frac{1}{2}M_5 + \frac{1}{30} M_8 -
\frac{1}{5}M_9,\\
P_{yy}&=  \frac{2}{3} M_1  - \frac{1}{2}M_5 + \frac{1}{30} M_8 - \frac{1}{5}M_9.
\end{split}
\end{align}
The time evolution equations for the set of  moments are:
\begin{eqnarray}
\label{mom}
\begin{aligned}
\partial_t M_1 + \partial_x M_2 + \partial_y M_3 &=0,\\
\partial_t M_2 + \partial_x \left(
\frac{2}{3} M_1  + \frac{1}{2}M_5 + \frac{1}{30} M_8 - \frac{1}{5}M_9\right)+ 
\partial_y M_{4} &=0,\\
\partial_t M_3  + \partial_x M_{4}+ \partial_y
\left(
\frac{2}{3}M_1  - \frac{1}{2}M_5 + \frac{1}{30} M_8 - \frac{1}{5}M_9\right) &=0\\
\partial_t M_{4} + \frac{1}{3}\partial_x \left(2 M_3 +
  M_7\right) + \frac{1}{3} \partial_y\left(2 M_2 + M_6\right)
&= \frac{1}{\tau} \left(M_{4}^{\rm eq}(M_1,M_2,M_3)- M_4\right),\\
\partial_t M_5+
\frac{1}{3}\partial_x \left(M_2  -M_6\right) +
 \frac{1}{3}\partial_y \left(M_7  -M_3\right)  &=
\frac{1}{\tau} \left(M_{5}^{\rm eq}(M_1,M_2,M_3) - M_5\right) ,\\
\partial_t M_6 -\frac{1}{5}
\partial_x\left(5 M_5 - M_8  + M_9 \right)
+\partial_y M_4 
&=\frac{1}{\tau}
\left(M_{6}^{\rm eq}(M_1,M_2,M_3) - M_6\right) ,\\
\partial_t M_7+ \partial_x M_4+\frac{1}{5}
\partial_y\left(5 M_5 + M_8  - M_9 \right) &= 
\frac{1}{\tau} \left(M_{7}^{\rm eq}(M_1,M_2,M_3) - M_7\right),\\
\partial_t M_8 + \partial_x\left( M_2 + 3 M_6\right)
+ \partial_y\left( M_3 + 3 M_7\right)
&= \frac{1}{\tau} \left(M_{8}^{\rm eq}(M_1,M_2,M_3) - M_8\right),\\
\partial_t M_9 
-\frac{1}{3}\partial_x\left(2 M_2 + M_6\right)- \frac{1}{3}\partial_y\left(2 M_3 + M_7\right)
&= \frac{1}{\tau} \left(M_{9}^{\rm eq}(M_1,M_2,M_3) - M_9\right).
\end{aligned}
\end{eqnarray}

The expression for the local  
equilibrium  moments $M^{\rm eq}_i$, $i=4,\dots,9$ in terms of the
basic variables $M_1$, $M_2$, and $M_3$ to  the order $u^2$ is: 
\begin{eqnarray}
\label{eqmom}
\begin{aligned}
M_4^{\rm eq}(M_1,M_2,M_3) &= \frac{M_2\,M_3}{M_1},\\
M_5^{\rm eq} (M_1,M_2,M_3)&=   \frac{M_2^2 - M_3^2}{M_1},\\
M_6^{\rm eq}(M_1,M_2,M_3) &= - M_2,\\
M_7^{\rm eq}(M_1,M_2,M_3) & =  - M_3,\\
M_8^{\rm eq} (M_1,M_2,M_3)&=  - 3\frac{M_2^2 + M_3^2}{M_1},\\
M_9^{\rm eq}(M_1,M_2,M_3) &= \frac{5}{3} M_1 - \frac{3\left(M_2^2 + M_3^2\right)}{M_1}.
\end{aligned}
\end{eqnarray}
The incompressible Navier-Stokes equations are
 the hydrodynamic limit of the system 
(\ref{mom}) and (\ref{eqmom}). 

In the next section, we shall remove small scales 
through a filtering procedure on the moment system (\ref{mom}), (\ref{eqmom}). 
A precise definition of the
small scales  is postponed until  later sections. For the
time being,  let us assume that there exist a  length-scale $\Delta$, and we
wish to look at the hydrodynamics at length-scale larger than $\Delta$
only.   
\section{Filtered kinetic theory}
\label{FKT}
Coarse-grained versions of the  Boltzmann equations have
been discussed in the recent literature  \cite{CSO,SKCO,JSP}.
However, a systematic treatment  is still lacking.
In this section, we shall fill this gap.
\subsection{Gaussian filter}
For any function $X$,  the filtered function $\overline{X}$ is defined 
as:
\begin{equation}
\label{filtering}
\overline{X}(\bx) = \int_{R^D} G(\br) X(\bx-\br) d \br.
\end{equation}
Function $G$ is called the filter. 
In the sequel, we apply the filtering operation (\ref{filtering})
on the moment system (\ref{mom}). We will need two relations.
First, for any function $X$,

\begin{equation}
\label{fprop}
\overline{\partial_{\alpha}X}=\partial_{\alpha}\overline{X}.
\end{equation}
This relation is sufficient to filter the propagation 
terms in the equation (\ref{mom}) due to  
linearity of propagation in the kinetic picture.
The latter  is a useful property which is {\it not} shared by
the hydrodynamic Navier-Stokes equations, where the nonlinearity
and nonlocality both come into the same ($\uu \nabla \uu$) term.
Any isotropic filter, which satisfies the condition of commuting of
the derivatives under the application of the filter (Eq. \ref{fprop}),
will suffice for the present purpose. 
We choose a standard 
Gaussian filter \cite{Pope2000} which has the property (\ref{fprop}):
\begin{equation}
 G(\br, \Delta) = \left( \frac{6}{\pi \Delta^2} \right)^D \exp{\left(
-\frac{6 \br^2}{ \Delta^2} \right)}.
\end{equation}
Let us recall the isotropy properties of a Gaussian filter:
\begin{eqnarray}
\label{isotropy}
\begin{aligned}
 \int_{R^D} G(\br, \Delta) d \br  &= 1,\\
\int_{R^D} G(\br, \Delta)  \br d  \br &= 0, \\
 \int_{R^D}  G(\br, \Delta) r_{\alpha} r_{\beta} d \br  &=
 \frac{\Delta^2}{12}\delta_{\alpha\beta}.
\end{aligned}
\end{eqnarray}

Second, in order to filter the nonlinear terms (\ref{eqmom}) 
in the right hand side of moment 
equations (\ref{mom}), we will  also need 
the following relation for
three arbitrary functions $X$, $Y$, $Z$ which follow immediately from the 
isotropy property by second-order Taylor expansion:
\begin{align}
\label{fnonlin}
\begin{split}
\overline{\left(\frac{XY}{Z}\right)} &=
\left(\frac{\overline{X}\  \overline{Y} }{\overline{Z}}\right) +
   \frac{\Delta^2}{12\, \overline{Z}} 
\Biggl\{ 
( \partial_{\alpha} \overline{X} )(\partial_{\alpha}
  \overline{Y})-\frac{2}{\overline{Z}} (\partial_{\alpha} \overline{Z})
\left( \overline{X} \partial_{\alpha} \overline{Y}
+\overline{Y} \partial_{\alpha} \overline{X}+
\frac{2\overline{X} \overline{Y}}{\overline{Z}}
\partial_{\alpha}
  \overline{Z}
\right) 
 \Biggr\}\\
&+ O(\Delta^4).
\end{split}
\end{align}
The effect of a Gaussian filter need not be truncated to any
order at the present step. The higher-order 
terms lumped under $O(\Delta^4)$ in equation (\ref{fnonlin}) can be computed from elementary
Gaussian integrals.  As we shall see it soon, higher than second order terms 
disappear in the hydrodynamic limit once the scaling of the filter-width versus Knudsen number
is appropriately chosen.

In the next section, we shall filter the moment equations (\ref{mom}). 

\subsection{Filtering the moment system}
Applying the filter (\ref{filtering}) to the moment system (\ref{mom}), (\ref{eqmom}),
using (\ref{fprop}) and (\ref{fnonlin}), and keeping terms up to the order $u^2$, 
we obtain the following filtered moment system: 
\begin{eqnarray}
\label{fmom}
\begin{aligned}
\partial_t \overline{M}_1 + \partial_x \overline{M}_2 + \partial_y
\overline{M}_3 =&0,\\
\partial_t \overline{M}_2 + \partial_x \left(
\frac{2}{3} \overline{M}_1  + \frac{1}{2}\overline{M}_5 + \frac{1}{30}
\overline{M}_8 - \frac{1}{5}\overline{M}_9\right)+ \partial_y
\overline{M}_{4} &=0,\\
\partial_t \overline{M}_3  + \partial_x \overline{M}_{4}+\partial_y
\left(
\frac{2}{3}\overline{M}_1  - \frac{1}{2}\overline{M}_5 + \frac{1}{30}
\overline{M}_8 - \frac{1}{5}\overline{M}_9\right) =&0,\\
\partial_t \overline{M}_{4} + \frac{1}{3}\partial_x \left(2 \overline{M}_3 +
  \overline{M}_7\right) + \frac{1}{3} \partial_y\left(2 \overline{M}_2 + \overline{M}_6\right)
=& \frac{1}{\tau} 
\left(M_4^{\rm eq}(\overline{M}_1,\overline{M}_2,\overline{M}_3) - \overline{M}_4\right)\\
+ \frac{\Delta^2}{12\,\tau\, \overline{M}_1} 
 (\partial_{\alpha}\overline{M}_2)(\partial_{\alpha}
 \overline{M}_3)
  + O\left(\frac{\Delta^4}{\tau}\right),
\\
\partial_t \overline{M}_5+
\frac{1}{3}\partial_x \left(\overline{M}_2  -\overline{M}_6\right) +
 \frac{1}{3}\partial_y \left(\overline{M}_7  -\overline{M}_3\right) 
 =&\frac{1}{\tau} \left(M_{5}^{\rm eq}(\overline{M}_1,\overline{M}_2,\overline{M}_3) -
   \overline{M}_5\right)\\
+ \frac{\Delta^2}{12\,\tau\, \overline{M}_1} 
\Biggl\{ 
(\partial_{\alpha}\overline{M}_2)(\partial_{\alpha}\overline{M}_2)
-(\partial_{\alpha}\overline{M}_3)(\partial_{\alpha}\overline{M}_3)
 \Biggr\} + O\left(\frac{\Delta^4}{\tau}\right),\\
\partial_t \overline{M}_6 -\frac{1}{5}
\partial_x\left(5 \overline{M}_5 - \overline{M}_8  + \overline{M}_9 \right)
+\partial_y \overline{M}_4 
=&\frac{1}{\tau}
\left(M_{6}^{\rm eq}(\overline{M}_1,\overline{M}_2,\overline{M}_3) - \overline{M}_6\right)
\\
\partial_t \overline{M}_7+ \partial_x \overline{M}_4+\frac{1}{5}
\partial_y\left(5 \overline{M}_5 + \overline{M}_8  - \overline{M}_9
\right) =& \frac{1}{\tau} \left(M_{7}^{\rm eq}(\overline{M}_1,\overline{M}_2,\overline{M}_3) -
  \overline{M}_7\right),\\
\partial_t \overline{M}_8 + \partial_x\left( \overline{M}_2 + 3 \overline{M}_6\right)
+ \partial_y\left( \overline{M}_3 + 3 \overline{M}_7\right)
=& \frac{1}{\tau} \left(M_{8}^{\rm eq}(\overline{M}_1,\overline{M}_2,\overline{M}_3) - 
\overline{M}_8\right)\\
- 
\frac{3 \Delta^2}{12\,\tau\, \overline{M}_1} 
\Biggl\{ 
(\partial_{\alpha}\overline{M}_2)(\partial_{\alpha}\overline{M}_2)
+(\partial_{\alpha}\overline{M}_3)(\partial_{\alpha}\overline{M}_3)
 \Biggr\}+ O\left(\frac{\Delta^4}{\tau}\right),\\
\partial_t \overline{M}_9 - \frac{1}{3}\partial_x\left(2 \overline{M}_2 + \overline{M}_6\right)
-\frac{1}{3}\partial_y\left(2 \overline{M}_3 + \overline{M}_7\right)
=& \frac{1}{\tau} \left(M_{9}^{\rm eq}(\overline{M}_1,\overline{M}_2,\overline{M}_3) -
  \overline{M}_9\right)\\
- 
\frac{3 \Delta^2}{12\,\tau\, \overline{M}_1} 
\Biggl\{ 
(\partial_{\alpha}\overline{M}_2)(\partial_{\alpha}\overline{M}_2)
+(\partial_{\alpha}\overline{M}_3)(\partial_{\alpha}\overline{M}_3)
 \Biggr\}+  O\left(\frac{\Delta^4}{\tau}\right).
\end{aligned}
\end{eqnarray}
Thus, we are set up to derive the hydrodynamic equations as the appropriate
limit of the filtered kinetic system (\ref{fmom}). 
In passing, we note that different moments relax with different
effective relaxation time scales, because the subgrid terms
are not the same for all kinetic moments.

\section{Hydrodynamic limit of the filtered kinetic theory}
\label{CE}
In the kinetic equation we have a natural length scale set by Knudsen
 number ${\rm Kn}$ (\ref{Knudsen}). 
The Navier-Stokes dynamics is obtained in
 the limit  ${\rm Kn} \ll1$.  By filtering the kinetic equation we have introduced
 a new length scale as the size of the filter $\Delta$.   
The hydrodynamic equations  produced by the filtered kinetic equation will
 depend on how  $\Delta$ scales with the Knudsen number.
 In order to understand this issue, let  us  look at the filtered equation for
 one of the moments  (\ref{fmom}) in the non-dimensional form.
In order to do this, let us introduce scaled time and space variables,
\begin{align}
\begin{split}
\bx^{\prime} &= \frac{\bx}{L},\\
t^{\prime} &=  \frac{tc_{\rm s}}{L},
\end{split}
\end{align}
where $c_{\rm s}=1/\sqrt{3}$ for the present model. Let us 
also specify Knudsen number in terms of the relaxation time $\tau$:
\begin{equation}
{\rm Kn} =  \frac{ \nu}{L c_{\rm s}} \equiv \frac{ \tau c_{\rm s}}{L} ,
\end{equation}
where, $\nu$ is the kinematic viscosity, $\nu=\tau
c_{\rm s}^2$ in the present model. Then, for example, the filtered equation for
the 
moment $M_4$ reads:

\begin{eqnarray}
\label{samplefmom}
\begin{aligned}
\partial_{t^{\prime}} \overline{M}_{4} &+ \frac{1}{3\, c_s}\partial_{x^{\prime}} \left(2 \overline{M}_3 +
  \overline{M}_7\right) + \frac{1}{3\, c_s} \partial_{y^{\prime}}\left(2 \overline{M}_2 + 
\overline{M}_6\right)
= \frac{1}{{\rm Kn}} 
\left(M_{4}^{\rm eq}(\overline{M}_1,\overline{M}_2,\overline{M}_3)-
  \overline{M}_4\right)\\
&+ \frac{\Delta^2}{12\,{\rm Kn} L^2 \overline{M}_1} 
\Biggl\{ 
( \partial_{x^{\prime}}
\overline{M}_2)
(\partial_{x^{\prime}}
 \overline{M}_3)
+
( \partial_{y^{\prime}}
\overline{M}_2)
(\partial_{y^{\prime}}
 \overline{M}_3)
 \Biggr\} + O\left(\frac{\Delta^4}{L^4 {\rm Kn}}\right).
\end{aligned}
\end{eqnarray}

We see that the in absence of the
filter ($\Delta=0$), the usual situation of a singularly perturbed kinetic
equation is recovered (and this results in the Navier-Stokes equations
in the first-order Chapman-Enskog expansion). Let us consider the following three possibilities 
of dependence of $\Delta$ on ${\rm Kn}$:

\begin{itemize}

\item If $\Delta/L\sim {\rm Kn}^{0}$,  then we do not have a 
singularly perturbed equation in (\ref{samplefmom}) anymore. That is,
the filter  is too wide, and it affects the advection terms in the
hydrodynamic  equations.

\item If  $\Delta/L\sim {\rm Kn}$, then we do have a singularly 
perturbed system. However, the subgrid terms are of order ${\rm Kn}^2$, and
they do not show up in the order ${\rm Kn}$ hydrodynamic equation. In other words, the filter
is too narrow so that it does not affect hydrodynamic equations at
the viscous time scale.

\item Finally, there is only one possibility to set the scaling of filter-width
with ${\rm Kn}$ so that the system is singularly perturbed, and the 
subgrid terms of the order $\Delta^2$
contribute just at the viscous time scale. This situation happens if
\begin{align}
\label{Scaling4}
\frac{\Delta}{L} \sim \sqrt{{\rm Kn}}.
\end{align}
Note that, with the scaling (\ref{Scaling4}), 
all the higher-order terms (of the order $\Delta^4$ and
higher) become of the order ${\rm Kn}$ and higher, so that they {\it do not
contribute at the viscous time scale}.

\end{itemize}

Once the scaling of the filter-width (\ref{Scaling4}) is introduced into
the filtered moment equations (\ref{fmom}), the application of the Chapman-Enskog
method \cite{Chapman} becomes a routine. We write:
\begin{equation}
\label{tDer}
\partial_t =\,\partial_t^{(0)}+ {\rm Kn} \,\partial_t^{(1)} + O({\rm Kn}^2),
\end{equation}

and for $i=4,\dots, 9$:
\begin{equation}
\label{Momexp}
\overline{M}_i = M_i^{\rm eq}(\overline{M}_1,\overline{M}_2,\overline{M}_3) + {\rm Kn}
\overline{M}_i^{(1)} + O({\rm Kn}^2).
\end{equation}



The hydrodynamics equations at the the order $O(1)$ are the Euler equations:
\begin{eqnarray}
\label{Euler}
\begin{aligned}
\partial_t^{(0)} \overline{M}_1 &=- \partial_x \overline{M}_2 - \partial_y
\overline{M}_3,\\
\partial_t^{(0)} \overline{M}_2 &=- \partial_x \left(
\overline{M}_1 c_s^2 + \frac{\overline{M}_2 \, \overline{M}_2 }{\overline{M}_1}
\right)- \partial_y
\left(\frac{\overline{M}_2 \, \overline{M}_3 }{\overline{M}_1}\right),\\
\partial_t^{(0)} \overline{M}_3  &=- \partial_x \left(\frac{\overline{M}_2 \, \overline{M}_3 }{\overline{M}_1}\right)-\partial_y
\left(
\overline{M}_1 c_s^2 + \frac{\overline{M}_3 \, \overline{M}_3 }{\overline{M}_1}
\right).\\
\end{aligned}
\end{eqnarray}
Note that no subgrid terms appear at
this time scale in the hydrodynamic equations (\ref{Euler}). 
This  means that large  scale motion, even after
filtering, is dictated just by the conservation laws. 
Zero-order  time derivatives of the non-conserved moments are evaluated using the chain
rule:
\begin{equation}
\label{derM1}
\partial_t^{(0)} M_{i}^{\rm eq}(\overline{M}_1,\overline{M}_2,\overline{M}_3) = 
\frac{\partial M_{i}^{\rm eq}}{\partial \overline{M}_1}
\partial_t^{(0)} \overline{M}_1 + 
\frac{\partial 
 M_{i}^{\rm eq}
}{\partial \overline{M}_2}
\partial_t^{(0)} \overline{M}_2
+
\frac{\partial 
 M_{i}^{\rm eq}
}{\partial \overline{M}_3}
\partial_t^{(0)} \overline{M}_3.
\end{equation}

In particular, to the order $u^2$:
\begin{eqnarray}
\label{derM0}
\begin{aligned}
\partial_t^{(0)} M_{4}^{\rm
  eq}(\overline{M}_1,\overline{M}_2,\overline{M}_3) &=
0,
 \\
\partial_t^{(0)} M_{5}^{\rm
  eq}(\overline{M}_1,\overline{M}_2,\overline{M}_3) &=
0, \\
\partial_t^{(0)} M_{6}^{\rm
  eq}(\overline{M}_1,\overline{M}_2,\overline{M}_3) &=
\partial_x \left(
\overline{M}_1 c_s^2 + \frac{\overline{M}_2 \, \overline{M}_2 }{\overline{M}_1}
\right)+ \partial_y
\left(\frac{\overline{M}_2 \, \overline{M}_3 }{\overline{M}_1}\right),
 \\
\partial_t^{(0)} M_{7}^{\rm
  eq}(\overline{M}_1,\overline{M}_2,\overline{M}_3) &=
 \partial_x \left(\frac{\overline{M}_2 \, \overline{M}_3 }{\overline{M}_1}\right)+\partial_y
\left(
\overline{M}_1 c_s^2 + \frac{\overline{M}_3 \, \overline{M}_3 }{\overline{M}_1}
\right),
\\
\partial_t^{(0)} M_{8}^{\rm
  eq}(\overline{M}_1,\overline{M}_2,\overline{M}_3) &=
0, \\
\partial_t^{(0)} M_{9}^{\rm
  eq}(\overline{M}_1,\overline{M}_2,\overline{M}_3) &=
-\frac{5}{3}\left[ \partial_x \overline{M}_2 + \partial_y
\overline{M}_3\right]
. 
\end{aligned}
\end{eqnarray}
At the next order $O({\rm Kn})$, correction to locally conserved moments is equal to zero,
\[ \overline{M}_1^{(1)}= \overline{M}_2^{(1)}=\overline{M}_3^{(1)}=0, \]
whereas  corrections to the non-conserved
moments, $\overline{M}_i^{(1)}$, $i=4,\dots,9$, are obtained 
by  substituting  Eq. \eqref{tDer}, and Eq. \eqref{Momexp} in Eq.
\eqref{fmom} and eliminating the zeroth order time derivatives using 
 Eq. \eqref{derM0}:
\begin{eqnarray}
\label{OknMom}
\begin{aligned}
  \overline{M}_{4}^{(1)} &= - L c_s 
\left[ \partial_x \overline{M}_3 +  \partial_y \overline{M}_2\right]
+ \frac{L^2 \,  }{12\, \overline{M}_1} 
\Biggl\{ 
({\partial_x\overline{M}_2}) \;
({\partial_x\overline{M}_3})
+
({\partial_y \overline{M}_2})
({\partial_y \overline{M}_3})
 \Biggr\},\\
  \overline{M}_{5}^{(1)} 
&= - 2 \, L c_s
\left[ \partial_x \overline{M}_2 -  \partial_y \overline{M}_3\right]
+ \frac{L^2 \,  }{12 \, \overline{M}_1} 
\Biggl\{ 
\left(
{\partial_x \overline{M}_2}
\right)^2
-
\left(
{\partial_x \overline{M}_3}
\right)^2
+\left(
{\partial_y \overline{M}_2}
\right)^2
-
\left(
{\partial_y \overline{M}_3}
\right)^2
 \Biggr\},
\\
 \overline{M}_{6}^{(1)} &=-
L c_s\left[
\partial_x \left\{\frac{3(\overline{M}_3^2)}{\overline{M}_1}
\right\}
+
 \partial_y \frac{6\,\overline{M}_2 \, \overline{M}_3}
 {\overline{M}_1}\right],
\\
  \overline{M}_{7}^{(1)} &=
-Lc_s\left[
 \partial_x \frac{6\,\overline{M}_2 \, \overline{M}_3}  {\overline{M}_1}+
\partial_y \left\{ \frac{3(\overline{M}_2^2)}{\overline{M}_1}
\right\}
\right]
,\\
  \overline{M}_{8}^{(1)} &=
{6\, L c_s}\left[
 \partial_x \overline{M}_2  +
\partial_y  \overline{M}_3
\right]- 
\frac{3 L^2 \,  }{12\ \overline{M}_1} 
\Biggl\{ 
\left(
{\partial_x \overline{M}_2}
\right)^2
+
\left(
{\partial_y \overline{M}_2}
\right)^2
+\left(
{\partial_x \overline{M}_3}
\right)^2
+
\left(
{\partial_y \overline{M}_3}
\right)^2
 \Biggr\},\\
  \overline{M}_{9}^{(1)} &=
6\, L c_s\left[
 \partial_x  \overline{M}_2  +
\partial_y  \overline{M}_3 
\right]- 
\frac{3 L^2 \,  }{12\, \overline{M}_1} 
\Biggl\{ 
\left(
{\partial_x \overline{M}_2}
\right)^2
+
\left(
{\partial_y \overline{M}_2}
\right)^2
+\left(
{\partial_x \overline{M}_3}
\right)^2
+
\left(
{\partial_y \overline{M}_3}
\right)^2
 \Biggr\}.
\end{aligned}
\end{eqnarray}
and the first-order time derivative of the conserved moments are:
\begin{eqnarray}
\label{OknEq}
\begin{aligned}
\partial_t^{(1)} \overline{M}_1 &=0,\\
\partial_t^{(1)} \overline{M}_2 &=- \partial_x \left(
\frac{1}{2}\overline{M}_5^{(1)}  + \frac{1}{30}\overline{M}_8^{(1)} -
 \frac{1}{5}\overline{M}_9^{(1)}  
\right)- \partial_y M_4^{(1)},\\
\partial_t^{(1)} \overline{M}_3  &=- \partial_x M_4^{(1)}
-\partial_y
\left(
-\frac{1}{2}\overline{M}_5^{(1)}  + \frac{1}{30}\overline{M}_8^{(1)} -
 \frac{1}{5}\overline{M}_9^{(1)}  
\right).\\
\end{aligned}
\end{eqnarray}
These equations shows that the viscous term and the subgrid term {\it both}
appear as the  $O(Kn)$ contribution. We remind  that
 no  assumption was made  about relative magnitude of the subgrid term as
compared with the viscous terms. 
The only requirement that is set on the subgrid scale term is
that they appear at the viscous time scale rather than the time scale
of the advection.  
 We can write the 
complete hydrodynamics equation by using Eq. \eqref{tDer}, Eq. \eqref{Momexp},
Eq. \eqref{Euler}, Eq. \eqref{OknMom}, and Eq. \eqref{OknEq} to obtain the
hydrodynamics equations correct up to the order $O({\rm Kn}^2)$ 
(at this stage one  recovers the
Navier-Stokes equations using the unfiltered kinetic equation). 
In the next section, we shall see how the subgrid
scale terms affect the Navier-Stokes description.

\section{Hydrodynamic equations}
The final set of hydrodynamics equation, valid up to the order $O({\rm
  Kn}^2)$ is:

\begin{eqnarray}
\begin{aligned}
\partial_t  \overline{\rho} + \partial_x   \overline{\rho u_x} + \partial_y
  \overline{\rho u_y}  &=0\\
\partial_t \overline{\rho u_x}  + \partial_x  \overline{P}_{x x} + \partial_y
\overline{P}_{x y} &=0\\
\partial_t   \overline{\rho u_y}  + \partial_x  \overline{P}_{x y} + \partial_y
\overline{P}_{y y}
 &=0,
\end{aligned}
\end{eqnarray}
where

\begin{eqnarray}
\begin{aligned}
\overline{P}_{xx} &= \left[\overline{P} + 
\frac{\overline{\rho u_x}^2}{\overline{\rho}} -2 \nu \overline{\rho}
    \overline{S}_{xx}
\right] + P^{\rm SG}_{xx},\\
\overline{P}_{xy} &= \left[\frac{
\overline{\rho u_x}\;
\overline{\rho u_y}}{\overline{\rho}} - 2 \nu \overline{\rho} \overline{S}_{x y}
\right] + P^{\rm SG}_{xy},\\
\overline{P}_{yy} &=\left[\overline{P}+ \frac{\overline{\rho
      u_y}^2}{\overline{\rho}} -2 \nu \overline{\rho}
    \overline{S}_{yy} 
\right] + P^{\rm SG}_{yy},
\end{aligned}
\end{eqnarray}

with   $\overline{P}=\overline{\rho} c_s^2$, as the  
 thermodynamic pressure  and 
\begin{eqnarray}
\nonumber
P^{\rm SG}_{\alpha \beta }   =  \frac{ {\rm Kn} L^2 \overline{\rho} }{12 }
\left[ 
\overline{S}_{\alpha\gamma} -\overline{\Omega}_{\alpha\gamma}
\right]\left[ 
\overline{S}_{\gamma\beta} +\overline{\Omega}_{\gamma\beta}
\right]
=   \frac{{ \nu }\, \overline{\rho}\, L}{12 \,c_s} \left[ 
\overline{S}_{\alpha\gamma} -\overline{\Omega}_{\alpha\gamma}
\right]\left[ 
\overline{S}_{\gamma\beta} +\overline{\Omega}_{\gamma\beta}
\right].
\end{eqnarray}

Thus, we have obtained a closed set of hydrodynamics equations, 
for appropriate
choice of filtering width. This set of equations, written in the
nondimensional form up to the order, $u^2$ is:
\begin{equation}
\label{incomp}
 \partial_{\alpha} \overline{ u_\alpha} =0,
\end{equation}
\begin{eqnarray}
\label{result}
  \begin{aligned}
\partial_t \,\left(\overline{ u_{\alpha}} \right) +\partial_\beta
\,\left(
{\overline{ u_{\alpha}}\; \overline{
    u_{\beta}}}\right) =& -
\partial_{\alpha}  P + 2 \,{\rm Kn}\, \partial_{\beta}
\left(
\overline{S}_{\alpha\beta} 
\right)
\\
&\;
-\frac{ {\rm Kn}   }{12 } \partial_{\beta} \left\{
\left(
\overline{S}_{\alpha\gamma} -\overline{\Omega}_{\alpha\gamma}
\right)\left( 
\overline{S}_{\gamma\beta} +\overline{\Omega}_{\gamma\beta}
\right)\right\}.
\end{aligned}
 \end{eqnarray}
Note that  the pressure appearing in the  momentum equation is not the
thermodynamic pressure anymore, but needs to be computed from the
incompressibility condition (Eq. \ref{incomp}) \cite{Majda}.  
Significance of the Knudsen number appearing in the equation (\ref{result})
is explained below in section \ref{END}.

Similar to the case of  the Navier-Stokes equation,  
the subgrid model (\ref{Smago1}) enjoys the consistent derivation from
the kinetic theory.
We  should remind here again that the result is independent 
of the particular kinetic model used for the derivation. Any  kinetic
model which recovers the Navier-Stokes equations in the hydrodynamic
limit will lead to the same result.


\section{Discussion and conclusion}
\label{END}
Now we shall summarise the results obtained in the present work and their
limitations:
\begin{itemize}
\item{It is possible to derive rigorously a coarse-grained closed set of equation
    for hydrodynamics, a long cherished goal in turbulence modeling.}
\item{The scale-separation present in the kinetic theory provides a
    natural way to obtain coarse models.}
\item{Arbitrary choice of filter-width is not allowed. }
\item{
In this work, we have shown that the operation of solving the
Boltzmann equation  
(Chapman-Enskog expansion) and coarse-graining
(filtering) do not commute. In the usual procedure of producing filtered
hydrodynamic equation, the filtering is done on the solution of the
Boltzmann equation  (Navier-Stokes equations), which leads to
closure problems. On the other hand,  reversing the order of these two 
operations, provides a physical meaning to the filtering width and
produces a closed set of equations in the hydrodynamic limit.}

\item{As the smallest length scale needed to be resolved in the new
    set of equations is $\Delta\sim {\rm Kn}^{1/2}\sim {\rm Re}^{-1/2} $, the
    cost of numerical computation reduces drastically as compared to
    the fully resolved simulation of the Navier-Stokes equations. We
    can get an estimate of this gain as follows:  The smallest scale needed
    to be resolved in the numerical simulation is proportional to the
    ${\rm Re}^{-1/2}$ rather than ${\rm Re}^{-3/4}$ (Kolmogorov scale). This changes  the
    scaling of number of degrees of freedom in a three-dimensional 
    simulation with ${\rm Re}$ from  ${\rm Re}^{9/4}$ to ${\rm
      Re}^{3/2}$ (number of
    grid point  is $ \propto
    \delta x^{-3}$, where $\delta x$ is the grid spacing). 
Further, a rough estimate of the scaling of the
    cost of time integration with ${\rm Re}$ is ${\rm Re}^{3/4}$ (number of
    time steps is $ \propto
    \delta x^{-1}$) in the case
    of fully resolved simulation of the Navier-Stokes equations 
    \cite{Pope2000}. However, in the present case this scaling will be
    ${\rm Ma}^{-1}$. This happens because any numerical scheme has to 
take a time-step dictated by the sound speed (or an analog of it pertinent to the discretization in time
chosen). Thus the scaling of the  
total cost of
    computation with the ${\rm Re}$ changes from ${\rm Re}^3$ for the
    Navier-Stokes equations to ${\rm Re}^{3/2}$ for the present equation. 
}
\item{ In the above estimation, the Mach number  ${\rm Ma}$ appearing in 
the equation for the
    filter-width  $\Delta$
    ($\Delta^{2} \propto {\rm Re}^{-1}$) was not  taken into the
    account. This is justified as long as ${\rm Re}$ is
    sufficiently large and the Mach number  ${\rm Ma}$ is not zero. An
    acceptable limit for the incompressible limit of the Navier-Stokes
    equation is ${\rm Ma} \sim 0.1$ (for example, most of the
    lattice-Boltzmann simulations of the incompressible Navier-Stokes
    uses ${\rm Ma} \sim 0.05-0.1$). Let us consider the case when the
    number of grid points in each direction is $1024$, then a fully
    resolved simulation of the Navier-Stokes equations is possible by
    taking the  Reynolds number as 
    ${\rm Re}\sim O(10^4)$, while a fully resolved simulation using the
    present subgrid model is possible by
    taking the Reynolds number as 
    ${\rm Re}\sim O(10^5)$, for ${\rm Ma}\sim 0.07$.    
    }
\item{  One  interpretation
of the subgrid scale terms is that the removal of the small-scales in
the kinetic picture appears naturally as the force term. As the extra
terms appearing in the evolution equations for non-conserved variables can 
also be generated in the unfiltered kinetic equation by an
appropriate choice of the external force field (dependent on the position
as well as molecular velocity).  Thus at least
formally, we can find a force-field which will act like a filter and
remove the small scales of motion present in the kinetic equation.
Thus the physical meaning of the filtering (a purely mathematical
operation) at the kinetic level is the application of some
self-consistent mean-field force which removes the small scale  of the
motions from the kinetic equation.  
   The technical
    advantage of the search for a mean-field force (appearing in the
    filtered kinetic equation), rather than an effective viscosity
    term (attempts to absorb subgrid scale contributions  in the
    viscous term of the Navier-Stokes equations) is that the one does
    not have to deal with the difficult question of what to do with
    the nonlinearity and nonlocality present in the convective term
    of the Navier-Stokes equations. 
}
\end{itemize}
 Finally, let us mention some further possible directions of study:

 \begin{itemize}
\item From a practical standpoint, a major goal  of going beyond a 
Navier-Stokes-based
coarse graining, is to make  a (filtered) kinetic theory 
work at possibly large ratios $\Delta/L$ . Indeed, successful
kinetic subgrid models have been known empirically for some time
\cite{SCI}. However, it is unclear why kinetic subgrid scale models
work better than models of the Navier-Stokes equations (for example, a 
recent comment on the kinetic models  of
turbulence was: ``Whether the approach can be supported by rigorous
theory remains to be shown'' \cite{BenziS}). 
Our analysis shows that everything is eventually ruled by 
the  smallness of the  Knudsen number, a well defined smallness
parameter present in the kinetic theory. This is the first rigorous
step in
the kinetic modeling of the turbulence.  
For example, the choice of the filter-width  (\ref{Scaling1}), based
on the   integer-power (standard) Chapman-Enskog analysis is a
conservative estimate only. 
In order to achieve a subgrid model {\it between} the
advection time scale (${\rm Kn}^0$) and the viscosity time scale (${\rm Kn}^1$),
that is, $\Delta/L\sim{\rm Kn}^{\gamma}$ with $0<\gamma<1/2$ requires 
a generalization of the Chapman-Enskog method to noninteger series
expansion in Knudsen number. This interesting possibility needs to be studied
separately.  The application of the method of the
invariant manifold \cite{GK94b}, which does not require Knudsen number
${\rm Kn}$ to be small, on the filtered kinetic equation for
$0<\gamma<1/2$  is a possible extensions of the present work.
The possibility of doing exact Chapman-Enskog expansion \cite{KG203,Slem},
also need to be investigated  further. Another  possible extensions is the use of
``renormalization-group'' ideas  by applying several filters of
increasing filter-widths.  

\item When the discrete-velocity kinetic theory of section (\ref{KT}) is appropriately
discretized in time and space, one arrives at the so-called entropic lattice Boltzmann method
\cite{KFOe99,AKOe2003}
(ELBM). In the ELBM, the thermodynamic stability (Boltzmann's $H$-theorem) is maintained
by the discrete-time $H$-theorem \cite{KGSB98,AK2002,SKC2002,Boghosian2003} which
results in unconditionally stable simulation algorithm for hydrodynamics.
It was argued \cite{AK2002} that ELBM is a built-in subgrid model.
It would be interesting therefore to establish a closer relation between ELBM and the
present theory.

\item Filtering the kinetic equations as above can be applied to a wide class of 
kinetic theories with a well-defined separation of time scales 
enrich existing resolved macroscopic models with physically sound
subgrid contribution (for example, the kinetic equations for the
granular flows \cite{kumaran}). 

\end{itemize}

 To conclude, the  presented coarse-grained equations are the first
 rigorously derived subgrid model.  Effectiveness
 of the model   in
 practice needs to be
investigated further numerically.  Work in this direction is currently in progress.
\section{\label{ack} Acknowledgments} 
I.\ V.\ K. and S.\ A. are 
 indebted to  A.\ N.\ Gorban for many  enlightening discussions, and suggestion
to lift macroscopic equations to a  kinetic theory before coarse-graining.
Useful discussions with K.\ Boulouchos, H.\ Chen, C.\ Frouzakis, S.\ Orszag, H.\ C.\
\"{O}ttinger, R.\ Thaokar,  and A.\ Tomboulides are 
kindly acknowledged.

\end{document}